\documentclass[a4paper,12pt]{article}
\usepackage{epsfig}
\usepackage[dvips,usenames]{color}
\usepackage{axodraw}
\usepackage{graphicx}

\newlength{\dinwidth}
\newlength{\dinmargin}
\date{}
\begin{document}
\title{  Radiative decays $J/\Psi\to\eta^{(\prime)}\gamma$ in perturbative QCD
}
\author{Ya-Dong Yang\thanks{E-mail address:
yangyd@henannu.edu.cn}\\
\small{Physics Department, Henan Normal University, Xin-Xiang,
Henan 453007,
P.R.China}\\
}
\maketitle
\begin{abstract}
With the recent investigations of
$g^{\ast}g^{\ast}-\eta^{(\prime)}$ transition form factor and
$\eta-\eta^{\prime}$ mixing scheme, we present an updated study of
the radiative decays $J/\Psi\to\eta^{(\prime)}\gamma$ in
perturbative QCD.  The decays are taken as a test ground for  the
$g^{\ast}g^{\ast}-\eta^{(\prime)}$ transition form factors and the
$\eta-\eta^{\prime}$ mixing scheme.   The form factors are found
to be working for glunic   $\eta^{(\prime)}$ productions and the
mixing angle is constrained to be
$\phi=35.1^{\circ}\pm0.8^{\circ}$.
\end{abstract}


\newpage

As it is well known that the heavy quarkonium  decays to light
mesons have played very important role in testing and
understanding QCD from the very beginning. The decays $J/\Psi \to
\eta^{(\prime)} \gamma$ are of great interests since they are
closely related to the issues of $\eta'-\eta$ mixing and $g^{*}
g^{*}-\eta^{(\prime)}$ transition form factors, which are very
important ingredients for understanding many interesting hadronic
phenomena of $\eta$ and $\eta^{\prime}$ productions. For example,
it would be very useful for explaining the large branching ratio
of strong penguin dominated decay $B\to K
\eta^{\prime}$\cite{cleoeta,belleeta, babareta}.

In the literature, studies of the decays $J/\Psi \to
\eta^{(\prime)} \gamma$ are different each other from the
treatments of formatting  gluons to $\eta^{(\prime)}$, namely,
direct nonperturbative $gg-\eta^{(\prime)}$ coupling through
strong anomaly\cite{Novikov} or two off-shell gluons coupled to
$\eta^{(\prime)}$ through quark loop\cite{korner}. In this letter,
we will take the second approach which had been pioneered
systematically within perturbative QCD by K\"orner, K\"uhn,
Krammer and Schneider(KKKS) \cite{korner} years ago.  In
Ref.\cite{korner}, the non-relativistic quark model and  the
weak-binding approximation were used for both heavy and light
mesons, and systematic helicity projectors were constructed to
reduce loop integrations. In this work, we follow their approach.
However, two improvements are included:
\begin{itemize}
    \item $g^{*}
g^{*}-\eta^{(\prime)}$ couplings are improved to be relativistic
transition form factors as advocated in Ref.\cite{ymz, Yang1, Ali}
in stead of non-relativistic modelling.
    \item The $\eta'-\eta$ mixing scheme is also updated to the
Feldmann-Kroll-Stech(FKS) mixing scheme\cite{FKS}.
\end{itemize}

 In  perturbative QCD  approach, the decays are depicted by the
 Feynman diagrams  in Fig.1.
To calculate the amplitudes for the decays, we need to know how to
deal with the dynamics of bound states. Generally, factorization
are employed.  Soft nonperturbative QCD bound state dynamics  are
factorized to the decay constants and the wave functions of
$J/\Psi$ and $\eta^{(\prime)}$ which will convolute  with the hard
kernel induced the decay. We shall use the non-relativistic
approximation for the heavy $J/\Psi$, but not for the  light
mesons $\eta'$ and $\eta$. Although a rigorous theory from first
principles for the light bound-states are still missing, some
effective approaches are in progress.  In recent years, it has
been realized that a proper treatment of the $\eta-\eta^{\prime}$
system requires a sharp distinction between the mixing states and
the mixing properties of the decay constants\cite{FKS}. Taking
strange-nonstrange flavor  basis for the $\eta-\eta^{\prime}$
system and the mixing of the decay constants following  the same
pattern of the state mixing, FKS have found a dramatic
simplification. They also have tested their mixing scheme against
experiment and determined corrections to the first order values of
the basic parameters from phenomenology.

\begin{figure}[htbp]
\begin{center}
\scalebox{0.9}{\epsfig{file=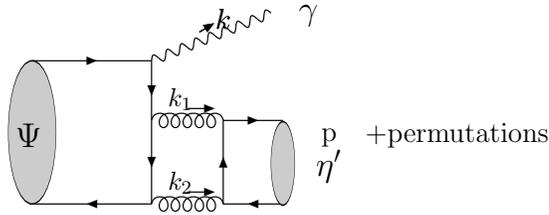}}
\end{center}
 \caption{\small  Lowest order QCD diagrams
 for $J/\Psi\to \eta^{(\prime)}\gamma$  decays.}
\end{figure}

In FKS mixing scheme the parton Fock state decomposition can be
expressed as
\begin{eqnarray}
 \mid \eta\rangle  &=& cos\phi \mid \eta_q\rangle -sin\phi \mid \eta_s\rangle,
\nonumber \\
 \mid \eta'\rangle &=& sin\phi \mid \eta_q\rangle +cos\phi \mid \eta_s\rangle
\end{eqnarray}
where $\phi$ is the mixing angle,  $|\eta_q\rangle \sim f_q
\Phi(x,\mu)|u\bar{u}+d\bar{d} \rangle/\sqrt{2}$ and
$|\eta_s\rangle \sim f_s \Phi(x,\mu)|s\bar{s}\rangle$. The decay
constants $f_q$, $f_s$ and the mixing angle $\phi$ have been
constrained from the available experimental data, $f_q=(1.07\pm
0.02)f_{\pi}$, $f_s=(1.34\pm 0.06)f_{\pi}$, $\phi=39.3^{\circ}\pm
1.0^{\circ}$ \cite{FKS}.

Already in Ref.\cite{Baier}, Baier and Grozin have derived
evolution equations for the distribution functions $\Phi(x,\mu)$
to the first order of $\alpha_s$, which
 eigenfunctions are found to be
\begin{equation}
\Phi(x, \mu)=6x(1-x)\left( 1+\sum_{n=2,4,...}^{} B_{n}(\mu)C^{\frac{3}{2}}(2x-1)
\right).
 \end{equation}
In the limit $\mu\to\infty$, the coefficients $B_{n}$ evolve to
zero and $\Phi(x,\mu)$ turns out to be $\phi_{AS}=6x(1-x)$. When
evolution equations  run down to low energy scale, its quark
contents mixed with glunic  states. However, the gluon content
enters the $\eta^{(\prime)}$ wave function from  next-to-leading
order. This observation encourages the calculations of  the $g^{*}
g^{*}-\eta^{(\prime)}$ transition
 form factors  similarly to the well known  $\gamma^{*}-\pi$ transition form factor
at leading order, which
 read \cite{ymz, Yang1, Ali}
\begin{eqnarray}
&&{\cal M}_{\mu\nu}=\langle
g^{\ast}_{a}g^{\ast}_{b}|\eta^{(\prime)}\rangle=-4\pi\alpha_{s}\delta_{ab}
i \epsilon_{\mu\nu\alpha\beta}Q_{1}^{\alpha}Q_{2}^{\beta}
F_{g^{\ast}g^{\ast}-\eta^{(\prime)}}(Q_{1}^{2},Q_{2}^2),\nonumber\\
&&F_{g^{\ast}g^{\ast}-\eta^{(\prime)}}(Q_{1}^{2},Q_{2}^2 )=
\frac{1}{2N_c }  f_{\eta^{(\prime)}}\int_{0}^{1}dx
\frac{ \phi_{\eta^{(\prime)}}(x,\mu)}{\bar{x}Q_{1}^2 +xQ_{2}^{2}
-x\bar{x}m^{2}_{\eta^{(\prime)}}+i\epsilon }
+(x\to \bar{x} ),
\label{formf}
\end{eqnarray}
where $\bar{x}=1-x$, $f_{\eta'}=\sqrt{2}f_{q}\sin\phi+f_{s}\cos\phi$ and
$f_{\eta}=\sqrt{2}f_{q}\cos\phi-f_{s}\sin\phi$.
 To the accuracy of this paper, $\phi_{\eta^{(\prime)}}(x,\mu)$
is taken to be the leading twist distribution functions(DAs)
$\phi^{AS}_{\eta^{(\prime)}}(x)=6x(1-x)$.

 Using $g^{*} g^{*}-\eta^{(\prime)}$ in Eq.\ref{formf} and  following the procedure
developed in Ref.\cite{korner, kuhn}, it is straightforward  to
evaluate the amplitudes for the decays as depicted by Feynman
diagrams in Fig.1. We get
\begin{equation}
 \Gamma(V\to \eta' \gamma)=\frac{1}{6}\left(\frac{2}{3}\right)^2 e_Q^2
\alpha^4_{s}(M_{V})
\alpha_e \frac{f^2_V  f^2_{\eta'} }{M^3_V } (1-z^2 ) {\mid}H(z){\mid}^2,
\end{equation}
where $z=m_{\eta'}/M_V$, $e_Q$ is  the heavy quark electric charge
and $\frac{2}{3}$ is the color factor. The dimensionless scalar
function $H(z)$ containing loop integrals is given by
\begin{equation}
H(z)=\frac{M^2_V }{2p{\cdot}k}\frac{1}{16}\frac{1}{i\pi^2 }\int^1_0 du
\phi^{AS}_{\eta'}(u)\int d^{4}q \frac{k_{1}{\cdot}k_2 (p{\cdot}kq^2-q{\cdot}k
q{\cdot}p)}
{D_1 D_2 k^2_1 k^2_2 (\bar{u}k^2_1 +u k^2_2 -u\bar{u} m^2_{\eta'} )},
\label{int1}
\end{equation}
where $D_1 =-k_{1}{\cdot}(k+k_2 )$, $D_2 =-k_{2}{\cdot}(k+k_1 )$,
$q=k_{1}-k_2$, and $p=k_{1}+k_{2}$.

 Obviously in Eq.\ref{int1}, the $k_{1}\cdot k_{2}$ numerator would cancel the $\eta'$
 form factor if it is taken be $\sim 1/k_{1}{\cdot}k_{2}$, and the hard scattering
kernel  would not convolute with the distribution functions of
$\eta^{\prime}$.

With the help of the algebraic identities
\begin{eqnarray}
&&q^2 =\frac{2}{M^2_V +m^2_{\eta'}}\left[ m^2_{\eta'}(D_{1}+D_{2})+M^2_V
(k^2_1 +k^2_2 )\right],\\
&& q{\cdot}p=k^2_1 +k^2_{2},~~~~ k_{1}{\cdot}k_{2}=\frac{1}{2}(p^2 -k^2_1 -k^2_2 )=
-\frac{1}{2}(p{\cdot}k+D_1 +D_2 ),
\end{eqnarray}
the integrand in $H(z)$ can be decomposed into a sum of four,
three and two-points functions which is presented in appendix A.
In the calculation of the loop integrals, we have used dimensional
regularization scheme and the methods developed in
Ref.\cite{hooft}

For numerical  results for the decays,  we  use
$\Gamma_{tot.}(J/\Psi)=(87\pm5)Kev$\cite{pdg}, $f_{J/\Psi}=400Mev$
and $\alpha_{s}(M_{J/\Psi})=0.2557$\cite{as}. We get
\begin{eqnarray}
{\cal B} ^{th}(J/\Psi\to\eta' \gamma)=3.9\times 10^{-3},&&
\left({\cal B} ^{exp}(J/\Psi\to\eta' \gamma)=(4.3\pm 0.3)\times 10^{-3},
PDG\cite{pdg}\right)\\
{\cal B} ^{th}(J/\Psi\to\eta \gamma)=3.5\times 10^{-4}, &&
\left({\cal B} ^{exp}(J/\Psi\to\eta \gamma)= (8.6\pm 0.8)\times
10^{-4}, PDG\cite{pdg}\right).
\end{eqnarray}
While ${\cal B} ^{th}(J/\Psi\to\eta' \gamma)$ agrees with
experiment, ${\cal B} ^{th}(J/\Psi\to\eta \gamma)$ turns out to be
too small. From the mixing scheme, it is easy to see that ${\cal B
} ^{th}(J/\Psi\to\eta' \gamma)$ is $insensitive$ to the mixing
angle $\phi$ when $\phi$ is about $35^{\circ}$, but ${\cal B}
^{th}(J/\Psi\to\eta \gamma)$ is very $sensitive$ to $\phi$. Take
$\phi=35.3^{\circ}$ fitted from $\eta'\to \rho\gamma$ and
$\rho\to\eta\gamma$ \cite{FKS}, we find
\begin{equation}
{\cal B} ^{th}(J/\Psi\to\eta' \gamma)=3.75\times 10^{-3},~~~~
{\cal B} ^{th}(J/\Psi\to\eta \gamma)=7.3\times 10^{-4},
\end{equation}
which  agree with experimental results quite  well. However, if we
take $\alpha_{s}(\mu)=\alpha_{s}{m_c}$, the results turn out to
overshoot their experimental data.

The most theoretical uncertainty may arise from the energy scale
choice in $\alpha_s (\mu) $. Because our calculation is performed
at the lowest order in QCD and there is no UV divergence in the
loop diagram induced the decay, we don't have strong argument to
choose a scale, as in usual case, to minimize the higher order
corrections by setting logarithm to zero. Naively, the scale could
be chosen from $m_c$ to $m_{J/\Psi}$. To reduce the scale
dependence, we relate ${\mathcal
B}(J/\Psi\to\eta^{(\prime)}\gamma)$ to ${\mathcal B}(J/\Psi \to
ggg)$
\begin{equation}
{\mathcal B}(J/\Psi\to\eta^{(\prime)}\gamma)= \frac{
\Gamma(J/\Psi\to\eta^{(\prime)}\gamma) }{\Gamma(J/\Psi\to ggg)}
{\mathcal B}(J/\Psi \to ggg). \label{bran}
\end{equation}
With the help of the known results\cite{ggg}
\begin{equation}
\frac{\Gamma(V\to ggg)}{\Gamma(V\to \mu^{+}\mu^{-})}
=\frac{10(\pi^2 -9)}{81 \pi e_Q^2 } \frac{\alpha^{3}_{s}(M)}
                 {\alpha_e^2 }
               \biggl\{ 1+\frac{\alpha_{s}(M)}{\pi} \biggl[ -19.4+
                 \frac{3}{2}\beta_0 \left( 1.16+\ln \left(
                 \frac{2M}{M_V}\right) \right) \biggr] \biggr\},
\end{equation}
we can get
\begin{equation}
{\mathcal B}(J/\Psi\to\eta^{(\prime)}\gamma)= \frac{9}{20(\pi^2
-9)} \frac{e_Q^2}{M_V^2} \frac{\alpha_{s}(M) \alpha_e
f^2_{\eta^{(\prime)}} (1-z^2 )|H(z)|^2 }{
1+\frac{\alpha_{s}(M)}{\pi} \biggl[ -19.4+
                 \frac{3}{2}\beta_0 \left( 1.16+\ln \left(
                 \frac{2M}{M_V}\right) \right) \biggr]}
                 {\mathcal B}(J/\Psi \to ggg).
\end{equation}
We will use the following relation and experimental data\cite{pdg}
for our numerical results
\begin{eqnarray}
{\mathcal B}(J/\Psi \to ggg)&=&{\mathcal B}(J/\Psi \to
hadrons)-{\mathcal B}(J/\Psi \to virtual\gamma\to hadrons)\nonumber\\
&=&(0.877\pm 0.005)-(0.17\pm 0.02) \nonumber\\
&=&0.707\pm 0.025.
\end{eqnarray}
Taking $\phi=35.3^{\circ}$ and $\alpha_{s}(M)=\alpha_{s}(m_{c})$,
we obtain
\begin{equation}
{\cal B} ^{th}(J/\Psi\to\eta' \gamma)=4.17\times 10^{-3},~~~~
{\cal B} ^{th}(J/\Psi\to\eta \gamma)=8.16\times 10^{-4},
\end{equation}
which agree with the experiment data.
\begin{figure}[htbp]
\begin{center}
\scalebox{0.9}{\epsfig{file=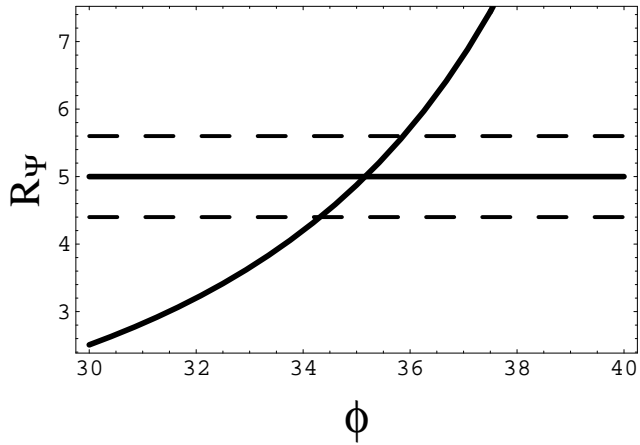}}
\end{center}
 \caption{\small  The ratio ${\cal R}_{J/\Psi}$ is shown by a solid
 curve as a function of $\phi$ (in degree).
The experimental data are shown by horizontal lines. The thicker
solid horizontal line is  its center value, thin horizonal dash
lines are its error bars.}
\end{figure}

 In Fig.2, we display the ratio
${\cal R}_{J/\Psi}={\cal B}(J/\Psi\to\eta' \gamma)/{\cal B}
(J/\Psi\to\eta \gamma)$
 as a
function of $\phi$, in which we expect that the relativistic and
the higher order QCD corrections may be concealed to large extent,
so that the ratio could be  predicted much more reliable than the
two decay rates respectively. Comparing our results with the
experimental measurement $R_{J/\Psi}=5.0\pm0.6$\cite{pdg} as
displayed by horizontal lines in Fig.2, we find
$\phi=35.1^{\circ}\pm0.8^{\circ}$ which is different from the
value $\phi=39.0^{0}\pm 1.6^{0}$ by $2\sigma$ also determined from
$J/\Psi\to \eta(\eta') \gamma$\cite{FKS} by using QCD anomaly
dominance mechanism formula. To make clear the origin of the
discrepancy between the two different determination of mixing
angle $\phi$, we recapitulate the key formula from the well known
work of Novikov $et~al.$ \cite{Novikov}
\begin{equation}
R_{J/\Psi}=\left| \frac{ \langle 0|G\tilde{G}|\eta'\rangle}
{\langle0|G\tilde{G}|\eta\rangle}\right|^2 \left(\frac{p_{\eta'}}
{p_{\eta}}\right)^{3}. \label{anomaly}
\end{equation}
This formula is frequently employed to determine $\eta-\eta'$
mixing angles in the literature.  Technologically, the strong
anomaly dominance  is  equivalent to the dominance of the ground
state and the neglect of continuum contribution to dispersion
relations as shown in Ref.\cite{Novikov,PBall}. So far,
considering the experimental and the theoretical uncertainties,
the difference between the predictions for ${\cal R}_{J/\Psi}$ by
the two mechanism is still marginal.  Although we have improved
$g^{\ast}g^{\ast}-\eta^{(\prime)}$ couplings  in Ref.\cite{korner}
from non-relativistic to relativistic, there is still large room
for theoretical improvements which is very worth for further
studying. We also note the CLEO/CESR-c project is going, where
about one billion $\Psi$ events would be produced. The  refined
measurements of these decays to be performed at CLEO-c will deepen
our understanding of the two $\eta^{(\prime)}$ production
mechanisms.

In this letter, we have studied the radiative decays
$J/\Psi\to\eta'(\eta)\gamma $ in perturbative QCD. The
relativistic $g^{\ast}g^{\ast}-\eta^{(\prime)}$ transition form
factors have been tested to be working  for $\eta^{\prime}$
production. The mixing angle in FKS scheme is constrained to be
$\phi=35.1^{\circ}\pm0.8^{\circ}$. This study  encourages further
applications of the form factor for  $\eta^{(\prime)}$ production
in hard processes. It is also very helpful for understanding the
abnormal large $\eta^{\prime}$ yields in B meson decays, which
have caught many theoretical attentions\cite{theta1, theta2}
recently.

\section*{Acknowledgments}
The author thank the members of the HEP group at Technion for
their hospitality where part of this work performed. Thanks to Gad
Eilam for his contributions to the early stage of this study. This
work is supported by the National Natural Science Foundation of
China under Contract No.10305003 and the Henan Provincial Science
Foundation for Prominent Young Scientists under Contract
No.0312001700.

\begin{center}
{\bf Appendix A}
\end{center}
In the evolution of the amplitudes for
$J/\Psi\to\gamma\eta^{(\prime)}$, we encounter the loop integral
in Eq.\ref{int1} which can be expanded in terms four, three and
two points functions
\begin{eqnarray}
H(z)=\frac{1}{16}\frac{1}{1-z^2 }\int du \phi_{AS}(u)
   \biggl[
        \frac{1-z^2}{2(1+z^{2})} 4^4 \left( m^4 D^{a}_{0}(u,z)-\frac{1}{2}
   (1-z^{2})m^4_V D^b_{0}(u,z)\right) \biggr. \nonumber \\
- \frac{1}{2}\left( 1-uz^2 \right)4^3 C^{b}_{0}(u,z)
-\frac{1}{2}\left( 1-2z^2 +uz^2 \right)4^3 C^{a}_{0}(u,z)
   \nonumber \\
  \biggl.
 +\frac{1}{2u} 4^2 \left(
B^a_{0}(u,z)-B^b_{0}(u,z)-B^c_{0}(u,z)+B^d_{0}(u,z) \right)
\biggr],
\end{eqnarray}
with the following functions
\begin{eqnarray}
D^a_{0}(u,z)&=&\frac{1}{8 m^4_{V} (1-u)uz^2 (1-z^2 )}
  \Biggl[ Sp\left( 1-\frac{1-\bar{u}z^2 }{\bar{u}(1-z^2 )}\right)
+2\pi i \ln \left( 1-\frac{1-\bar{u}z^2 }{\bar{u}(1-z^2 )}-i\epsilon\right)
\Biggr.
\nonumber \\
&+&Sp\left( 1-\frac{1-\bar{u}z^2 }{1-(1-2u)z^2 }\right)
+2\pi i \ln \left( 1-\frac{1-\bar{u}z^2 }{1-(1-2u)z^2 }-i\epsilon\right)
-Sp\left( 1-\frac{\bar{u}-(1-2u)z^2}{\bar{u}(1-z^2 )}\right)
\nonumber \\
\Biggl. &+&\left(2\pi i+\ln\left(
\frac{\bar{u}-(1-2u)z^2}{1-\bar{u}z^2}\right)\right) \left( \pi i
+\ln\left( \frac{(1-z^2 )(\bar{u}-(1-3u+2u^2 )z^2 ) } {u
z^2}\right)\right) \Biggr],
 \end{eqnarray}

\begin{eqnarray}
D^b_{0}(u,z)&=&\frac{1}{4 m^4_{V} (1-(1-2u)z^2 )u(1-z^2 )}
  \Biggl[ 2Sp\left( -\frac{1-z^2 }{uz^2 }\right)-
2Sp \left( -\frac{(1-2u)(1-z^2 )}{u}\right)
\Biggr.
\nonumber \\
&+&Sp\left( -\frac{1-z^2 }{z^2 (\bar{u}-(1-3u+2u^2 )z^2 )}\right)
-Sp \left( -\frac{(1-2u)^2 z^2 (1-z^2 )}{\bar{u}-(1-3u+2u^2 )z^2
}\right)
\nonumber \\
\Biggl.
&+& \ln\left( \frac{1-\bar{u}z^2}{z^2 (\bar{u}-(1-2u)z^2}) \right)
 \ln\left( \frac{(1-z^2 )(\bar{u}-(1-3u+2u^2 )z^2 ) }
{u z^2}+i\pi \right) \Biggr],
 \end{eqnarray}

\begin{eqnarray}
C^b_{0}(u,z)&=&-\frac{1}{m^2_V}\int^1_0 dy \frac{1}{
4u^2 z^2 -y^2 (1-2z^2 )-2uy(1-3z^2 ) -i \epsilon} \nonumber\\
            && \ln \left(
\frac{y(2(1-2z^2 )-y(1-2z^2 )-2u(1-3z^2))-i\epsilon}
{2(y(1-z^2 )-2u^2 z^2 )-i\epsilon }
\right), \\
C^a_{0}(u,z)&=&-\frac{1}{m^2_V}\int^1_0 dy \frac{1}{
y^2 +2y\bar{u}(1-z^2 )+4\bar{u} z^2  -i \epsilon} \nonumber\\
            && \ln \left(
\frac{-y(y+2\bar{u}(1-2z^2 )+i\epsilon}
{4\bar{u}^2 z^2 +i\epsilon }
\right),\\
 B^a_{0}(u,z)&=&\frac{2}{\epsilon}-\gamma_E +\ln4\pi+\ln\mu^2 +2-\ln m_V^2
 \nonumber \\
&&-\left(1-\frac{1}{(1-2u)(1-2uz^2 )+i\epsilon}\right)\ln
    \left(1-(1-2u)(1-2uz^2 )+i\epsilon \right),\\
B^b_{0}(u,z)&=&\frac{2}{\epsilon}-\gamma_E +\ln4\pi+\ln\mu^2 +2-\ln m_V^2 ,\\
 B^c_{0}(u,z)&=&\frac{2}{\epsilon}-\gamma_E +\ln4\pi+\ln\mu^2 +2-\ln m_V^2
 \nonumber \\
&&-\left(1-\frac{1}{(1-2u)(2\bar{u}z^2 -1 )+i\epsilon}\right)\ln
    \left[1-(1-2u)(2\bar{u}z^2 -1)+i \epsilon\right],\\
B^d_{0}(u,z)&=&\frac{2}{\epsilon}-\gamma_E +\ln4\pi+\ln\mu^2
+2-\ln m_V^2 -\frac{2(1-z^2 )}{1-2z^2 }\ln[ 2(1-z^2 )]+\ln(1-2 z^2
).
\end{eqnarray}

Where $Sp(x)= Li_{2}(x)$  is the Spence function and $\bar{u}=1-u$.

\end{document}